\documentstyle[a4]{article}
\newcommand{\er}{\end{eqnarray}}
\newcommand{\br}{\begin{eqnarray}}
\newcommand{\be}{\begin{equation}}
\newcommand{\ee}{\end{equation}}
\newcommand{\epe}{\end{equation}}
\newcommand{\bea}{\begin{eqnarray}}
\newcommand{\eea}{\end{eqnarray}}
\newcommand{\ba}{\begin{eqnarray}}
\newcommand{\ea}{\end{eqnarray}}
\newcommand{\epa}{\end{eqnarray}}

\newcommand{\dsl}
  {\kern.06em\hbox{\raise.15ex\hbox{$/$}\kern-.56em\hbox{$\partial$}}}

\def\a{\alpha}
\def\b{\beta}
\def\r{\rho}

\def\d{\delta}

\def\ep{\epsilon}

\newcommand{\ZZ}{{\rm \kern 0.275em Z \kern -0.92em Z}\;}
\begin{document}
\begin{center}
{\Large Lorentz Symmetry Breaking and Planar Effects from
Non-Linear Electrodynamics.}
\\
\vspace*{1.2cm} {\large
Marcelo Botta Cantcheff\footnote{e-mail: botta@cbpf.br}}

\vspace*{3mm} Instituto de F\'{\i}sica Te\'orica (IFT/UNESP)

Rua Pamplona, 145 - Bela Vista

01405-900 - S\~ao Paulo, SP - Brazil.

\vspace{3mm} Centro Brasileiro de Pesquisas Fisicas (CBPF)

Coordena\c c\~ao de F\'\i sica Experimental de Altas Energias (LAFEX)

Rua Dr. Xavier Sigaud, 150 - Urca

22290-180 - Rio de Janeiro - RJ - Brazil.

\end{center}

\begin{abstract}
\noindent

We propose a modification of standard linear
electrodynamics in four dimensions, where effective non-trivial
interactions of the electromagnetic field with itself and with
matter fields induce Lorentz violating Chern-Simons terms.
This yields two consequences: it provides a more
realistic and general scenario for the breakdown of Lorentz
symmetry in electromagnetism and it may explain
the effective behavior of the electromagnetic field in certain
planar phenomena (for instance, Hall effect). A number of
proposals for non-linear electrodynamics is discussed along the
paper. Important physical implications of the breaking of Lorentz
symmetry, such as optical birefringence and the possibility of having
conductance in the vacuum are commented on.

\end{abstract}

\section{Introduction.}

It is often claimed that the correct description of many
electromagnetic phenomena in three dimensions is provided by the
Maxwell-Chern-Simons theory \cite{MCS3d,winder}. It is believed to
describe electromagnetic interactions in very thin films. However,
if we accept that three-dimensionality is an effective circumstance
(i.e, the world is four dimensional, but the physics of a certain
phenomenon would be approximately described by a field theory on a
three-dimensional manifold), a fundamental question arises: what is
the mechanism that explains this behavior of the electromagnetic
field from the four-dimensional point of view?

In most physical situations modelled by a planar theory, where
a Chern-Simons term appears, the gauge field $A$ is
considered as an effective one, describing other collective, more
fundamental, degrees of freedom (fermionic or bosonic) in some
material medium \cite{zhang,fradkin}. However, they all suppose a
three-dimensional manifold as its starting point
\cite{zhang,fradkin,FQHE}. Actually, the appearance of this term is
badly motivated in planar models which effectively describe
electromagnetism in thin material films. Furthermore, a
dimensional reduction scheme from the ordinary Maxwell theory,
interacting with charged matter in the usual way,
is also hard to be argued. If we accept that at a macroscopic scale (i.e,
the scale of electromagnetic interactions) the space-time is effectively
four-dimensional, even in the context of electrodynamic planar phenomena,
any compactification scenario is not well justified as starting point.

A few years ago, a modification of Maxwell's electromagnetism in
four dimensions has been proposed which considers a kind of
Chern-Simons term in the action, $\int dx^4\, V_{\alpha}\,
\ep^{\alpha\beta \mu\nu}A_{\beta} F_{\mu \nu}$, where Lorentz
symmetry is explicitly broken by an external vector, $V^\mu$
\cite{jac}. There is a growing literature considering this proposal
seriously, and exploring the possible origin and consequences of
this approach \cite{consecuencias}.

In fact, in the present approach, we emphasize that broken Lorentz symmetry
(abbreviated as BLS) would be crucial to obtain a Chern-Simons term in some
three-dimensional effective model (since it is {\it automatically}
induced from a BLS term, as will be explained in Section 1 of this paper).
Reciprocally, the physical contexts for which an effective planar
model with a Chern-Simons term is more appropriate \cite{MCS3d}
{\it must dictate} precisely what are the ways in which BLS
should appear in a fundamental 4-dimensional electrodynamics. This
is the paradigm which, together with the assumption of Lorentz
invariance, motivates our construction. The model we
are going to build up here assumes a non-linear (but {\it
relativistic}) electrodynamics which induces a modification of
this kind (BLS) of the standard Maxwell theory. Furthermore, it
is shown that BLS does not need to be introduced {\it by hand},
but it can naturally appear in some realistic physical situations.
For example, in the presence of external fields and/or when non-uniform
distributions of charged matter are considered (and consequently,
a CS term will be induced in an embedded hyperplane).

There are additional theoretical motivations for our construction.
Recent works \cite{todos,paschoal} consider BLS with very interesting
consequences on certain aspects of condensed matter phenomena,
such as the proposal of a non-minimal coupling in order to describe the
quantum Hall effect. However, the vector $V$ is
identified by hand with the external magnetic field,
orthogonal to the Hall sample\footnote{And, in Ref.
\cite{todos}, BLS is used to describe the Pauli coupling \cite{pauli}.}.
In this paper we consider these observations
and accomodate them as natural consequences of a
more fundamental non-linear electrodynamics theory, in a general
and elegant way, aiming that such {\it ad hoc} interpretations are
not necessary.

%


%

This work is organized as follows: in Section 2, we define our
proposal in general, writing down the basic postulates for
nonlinear electrodynamics. In Section 3, we extend these
assumptions to theories in which the electromagnetic field interacts
with other charged ones. In Subsection 3.1, we define a non-linear
coupling to generic matter and argue on the appearance of BLS
terms and non-minimal couplings. In Section 4, we discuss the Hall
effect in terms of non-linear theories. In Section 5, we carry out
BLS with examples of non-linear theories. We present our final
observations in Section 6.

\section{Non-linear electromagnetism and BLS.}

It appears to be difficult to work with a theory whose form is
unknown. However, we propose here a technique that allows (under
a few simple and general assumptions) to evaluate the
effective action around certain specific background solutions {\it
without} the knowledge of the full non-linear theory.
The assumptions are as follows:

\begin{enumerate}

\item[(I)] Assume that electromagnetism is described by non-linear
equations. Thus, the Lagrangian $L_{NL}$ is a generic (in
principle, non-quadratic), relativistically invariant, functional
of a vector field $A_\mu$ (and $F_{\mu \nu}\equiv 2\partial_{[\mu}
A_{\nu]})$\footnote{We neglect the symmetric derivatives because
we only want to consider degrees of freedom corresponding to a spin-1
boson.}.

\item[(II)] There are non-trivial, stable solutions $ ( A^0_\mu \, ,F^0_{\mu \nu}) $
of the classical equations of motion, such that small
field perturbations (i.e $ \| F_{\mu \nu}- F^0_{\mu \nu}\|\ll \|F^0_{\mu \nu} \|
$ ) are properly described by the second order expansion of the
action about these background solutions: \be \label{1} S_L [ A_\mu
]= S[A^0] + \int dx dy (A_\mu - A^0_\mu )(x) \left(\left. \frac12  \frac{\d^2
S}{\d A_\mu (x)\d A_\nu (y)} \right|_{A^0} \right) (A_\nu - A^0_\nu )(y)
.\ee In particular $F^0_{\mu \nu} \equiv 0$ is a solution, and linearization
around this one coincides with the standard Maxwell theory.


Clearly, the quantization of this theory in the {\it saddle point}
approximation is contemplated in this statement because quantum
fluctuations are supposed to be much weaker than the classical
background fields. So, in order to quantize the electromagnetic field
and find out the photon spectrum, one considers the free theory (\ref{1}).


\item[(III)] The linearized theory around $F^0_{\mu \nu}=0$ has the usual $U(1)$-gauge
invariance $A_\mu \to A_\mu + \partial_\mu \a$.

This assumption is motivated in the original reference \cite{jac}
where gauge invariance of electrody\-na\-mics is strongly relaxed
(a mass term is considered within observable limits). Here we
admit this possibility, but we require that these limits depend of the
strength of the background field. So, when it is weak ($F^0_{\mu \nu} \to
0$), gauge invariance shall be recovered in order that the theory agrees with
the full Maxwell theory in ordinary physics.

\end{enumerate}


Now, we are going to show that these simple assumptions on the
unknown non-linear theory are sufficient to recover the behavior
that is relevant for us. The linearized effective action has the general form
\be \label{2-0}
S_L[ A_\mu ]= \frac{1}{4}\int d^4 x [C_1^{\a\b\mu\nu} F_{\a
\b}F_{\mu \nu} + C_2^{\a\mu\nu} A_{\a }F_{\mu \nu} + C_3^{\mu\nu}
A_\mu A_\nu ]   , \ee where the tensors $C_i $ do not depend
explicitly on the spacetime point, being only algebraic
functionals of the background fields, $F^0_{\a \b}, A^0_\mu$. The
crucial observation, over which our construction stands, is that
the tensors $C_i$ are typically non-trivial, which leads to BLS.

As a first example, a very simple gauge invariant non-linear
theory for pure electromagnetism is \be  \label{BLS2}
L_{\r}[A_{\mu}]\,= -{\b \over 4} \, F_{\mu \nu}F^{\mu \nu} +
\frac{1}{4} \, \r(F^2)\, \ep^{\alpha\beta \mu\nu}F_{\a\beta}
F_{\mu \nu} .\ee By expanding this theory about a non-constant
solution, $F^0_{\mu \nu}(x)$, we get a BLS theory where an
external vector $V_\mu$ appears, given by $V_\mu =
-\frac14 \partial_\mu \rho(F_0^2(x))$

Now, because we wish to discuss the consequences of this approach
in some specific realistic scenario with physical
interest, let us consider an example where the background solution
consists of an uniform magnetic field, $B_\mu \equiv
\epsilon_{\mu\nu\a\b}t^\nu
F^{0\,\a \b}$ (where $t^\nu = \partial^\nu t$ is a unit
timelike four-vector and $t$ is the time coordinate in the rest frame),
and the other components of $F_{\mu \nu}$ vanish. Then, clearly, $C_i[F^0]$
are algebraic functions of the vectors $B^\mu , t^\nu $,
the space-time metric tensor $ \eta_{\mu \nu}$, and the
Levi-Civita tensor $ \epsilon_{\mu\nu\a\b}$. In order to simplify the analysis
we assume that these tensors do not depend on $t^\nu$. So, by seeking all the
possible independent combinations, the most general form for them
reduces to: \bea \label{2} C_1^{\a\b\mu\nu} = - \b (B^2)
h^{\a\mu} h^{\b\nu}
\nonumber\\ C_2^{\a\mu\nu} = \rho(B^2)\epsilon^{\mu\nu\a\b}B_\b \nonumber\\
C_3^{\mu\nu} = m(B^2)  \eta^{\mu \nu} + k(B^2) B^\mu B^\nu \;
,\eea where, for convenience, we have expressed $h_{\mu\nu}=
\eta_{\mu\nu} - \kappa(B^2)\, B^{-2} B_\mu B_\nu $. Other possible
combinations in $C_1$ do not contribute due to the antisymmetry of
$F$, and many other ones (for instance $\epsilon^{\mu\nu\a\b}A_\a
\partial_\b F_{\mu \nu}$), were dropped out because they may be absorbed
into $C_2$ and/or $C_3$, or contribute to boundary terms (e.g
$\epsilon^{\mu\nu\a\b} F_{\mu \nu}F_{\a \b}$). There also appear
terms depending on $\partial_\mu A^\mu$ or $k_\mu A^\mu$ (for some
vector $k_\mu$) but they are not gauge invariant and could be
considered to be part of the gauge-fixing sector of the theory.

Gauge invariance of $S_L$ implies additional restrictions on these
general combinations: $C_3$ must be antisymmetric, and therefore
$C_3 = m = k =0$. Furthermore $\partial_\mu C_2^{\a\mu\nu}=0$
implies $\partial_{[\mu} (\r B)_{\nu]} =0$ which is satisfied for a
constant $B_\mu$. Finally, by virtue of assumption (II), the
coefficients must satisfy $\r(0)\in\Re \; ,\; h(0)=0$.

So, the resulting gauge invariant part of the Lagrangian
is precisely the Maxwell-CS theory in four dimensions \be \label{BLS}
L_{L}[A_{\mu}]\,= -{\b \over 4} \, F_{\mu \nu}F^{\mu \nu} +
\frac{1}{4} \, (\r B _{\alpha})\, \ep^{\alpha\beta
\mu\nu}A_{\beta} F_{\mu \nu} \ee proposed by Carroll, Field and
Jackiw \cite{jac}. This theory breaks Lorentz symmetry for the
external background vector $V_\mu = \r B_\mu$, which shows that
the identification speculated in \cite{paschoal} could be derived
in a framework of this type.

Let us briefly explain why the Chern-Simons action, {\it
automatically} appears in the BLS action when we search for planar
features (thus turning dimensional reduction unnecessary, as commented
in the introduction). In fact the BLS action is actually a CS theory in
$(2+1)$-dimensions embedded in $(3+1)$-dimensions, and by itself, it
does not encode any information on the field-dependence in the
direction of the external (for instance, space-like) vector $V$:
if $z$ is its affine parameter, i.e. $V=\frac{\partial}{\partial z
}$, then we get a foliation of the spacetime in
$(2+1)$-hypersurfaces $\Sigma_z$ parametrized by $z$ (and $V$ is
orthogonal to each hypersurface\footnote{Notice that if the
space-time (or the  space-time region considered in the
integration) is simply connected, the condition of existence of
this $z$-coordinate is {\it equivalent} to gauge invariance of the
action, namely $dV=0$.}). Therefore, the BLS action may be written as \be
S_{BLS} =
\int_0^L dz \, S_{CS}[A(z) , \Sigma_z], \ee where \be S_{CS}[A(z)
, \Sigma_z]=\int_{\Sigma_z} A(z) \wedge d A(z) \, ,\ee is the
Chern-Simons action for the 1-form gauge field $A(z)$ on a
three-dimensional manifold $\Sigma_z$. Thus, the dependence of
this field on the parameter $z$ is not determined by this theory.
It only has to satisfy usual convergence conditions. For example, if the
interval $(0,L)$ extends to $(-\infty, +\infty)$, $A(z)$ has to be an
square-integrable function ($A \in L^2(\Re)$). In this sense, we
can interpret the BLS action simply as a sum of Chern-Simons
theories on manifolds $\Sigma_z$.

Consequently, at the low-energy limit of the electromagnetic field,
the Maxwell term in (\ref{BLS}) is ne\-gli\-gible for the planar
dynamics and then we get a planar Chern-Simons theory {\it without}
any dimensional reduction (which would have to be further
justified\footnote{Alternatively, a dimensional reduction scheme may
be considered \cite{todos} when $V$ does not coincide with the
coordinate of reduction, but in this case, the part of the action
which depends on derivatives in the $V$-direction is lost.}). Notice
that this describes an eventual situation of confinement of the
electromagnetic field (photon) into a $(2+1)$-manifold, which does
not result from a constraint of the charged matter into a planar
sample.

Let us point out that away from the low energy regime (i.e, when
the Maxwell term contributes), there could also be a totally
planar phase, where the contribution of the term $V^\mu
F_{\mu \nu}$ to the action would become negligible. Since the
coefficients $\b, \theta, h, ..$ depend on the particular
non-linear theory, the condition for this is that $h_{\mu\nu}$ is
the flat 2+1-metric $h_{\mu\nu} = \eta_{\mu\nu} - B^{-2} B_\mu
B_\nu$ for some value of the background field. For instance, the
function $\kappa(B^2)$ could go to $1$ as $B^2 \to \infty$.
In this case we would not only have BLS, but {\it exact planar
Maxwell-Chern-Simons electromagnetism} in the presence of a strong
background field $B_\mu$. In Section 4 we argue on a planar
projection, in the context of the Hall effect.

It is interesting to notice that this BLS-term which, as shown here,
 is nothing but a genuine Chern-Simons term (in $(2+1)$ dimensions).
  It may also be obtained (see reference \cite {klink}) by considering
   the effective action of a standard (Lorentz and CPT invariant) gauge
    theory in four dimensional Minkowski space-time, such that a spatial
     dimension is taken to be compact ($\sim M \times S^1$). Then, by
      integrating out the chiral fermions while keeping gauge invariance,
       a Chern-Simons theory on the uncompactified part (three-dimensional) of spacetime $M$ appears.

This is remarkable because, despite what was said in
 the last paragraph, we {\it are not} considering non-trivial
  space-time topologies. This opens the possibility of obtaining
   such a term, alternatively, in a more standard electrodynamics
    theory in which the background would have non-trivial topology.
     However, as it happens in all approaches based on compactification
      of the space-time, it is not clear what is the physical reason behind this compactification.

\section{Non-linear electrodynamics: BLS in matter backgrounds.}

If we consider non-linear electrodynamics involving possible
non-linear couplings between charged fields and the
electromagnetic potential, BLS could also appear due to special
(background) distributions of charged matter (thin distributions,
for instance).

So, the possibility we are going to present here shall provide a
more realistic phenomenological context in which BLS could appear. Up
to now, BLS was most related to astrophysical situations
\cite{jac,astroV1,astroV2} or very high energy physics
\cite{kosteleky}, rather than to the presence of matter or
background fields in general. In the present approach, we wish to
introduce this point of view.

Thus, one must assume a total (non-linear) action for electrodynamics
\be S_{ED}=S[A,F,\bar{\psi}, \psi, \phi,
\phi_\mu]\, , \ee where $\bar{\psi}, \psi$ are fermionic charged
fields and $\phi, \phi_\mu$ describe bosonic matter fields with
spin-0 and -1, respectively. We will not make here any hypothesis
on the nature of these fields. However, one experimentally knows
that electromagnetism in the vacuum is accurately described by the
observed Maxwell theory. Thus, the only possibility of having some
deviation at some observable level is, perhaps, near or in the
interior of certain special distributions of matter (e.g. thin
films)\footnote{ Or in presence of external electromagnetic field
in a non-linear theory, as argued in the first section.}.
Therefore, these fields could be interpreted as effective ones,
describing perhaps, collective degrees of freedom associated to
the material medium. So, for example, these fields could be
describing phonons, which are believed to be coupled to the
electromagnetic field in material media, or some condensate of
fundamental charged fields (this is the case of the Cooper-pairs,
which should be important to describe the superconductivity
phenomena).

\vspace{0.7cm}

So, from now on, we assume that {\it the postulates}
(I, II, III) {\it extend to theories in which the electromagnetic
field interacts with other ones}.

\vspace{0.7cm}

Next, let us assume, for simplicity, a background
distribution of a charged scalar field, $\phi_0 (x)$. Following
the procedure established in Section 1, arguing the validity
of equation (\ref{1}), we also get tensors $C_i =
C_i(\phi_0,\partial_\mu \phi_0,
\partial_\mu \partial_\nu \phi_0, \dots)$ as in expression
(\ref{2}). We assumes, for simplicity, that the degrees of freedom
involved in the electromagnetic interaction are precisely the ones
corresponding to a standard charged spin-0 boson. Also, to avoid the
appearance of ghosts, second (or higher) order derivatives will not
be considered in the action. So, we get $C_i =
C_i(\phi_0,\partial_\mu \phi_0)$. In particular, the contribution to
BLS is given by $C_2$ (or equivalently $C_1$), which in general
reads: \be \label{C2} C_2^{\a\mu\nu} =
\epsilon^{\mu\nu\a\b}\partial_\b \rho(\phi_0) = \rho'(\phi_0)
\epsilon^{\mu\nu\a\b}\partial_\b \phi_0 \; \; \; \left( \Rightarrow
V_\mu = \nabla_\mu \r \right)\, .\ee Thus, the most general
linearized gauge invariant Lagrangian for electromagnetism in a
background $\phi_0$ is:

\be \label{Lphi0} L_{eff}[A_{\mu}]\,=
-{\b(\phi_0) \over 4} F_{\mu \nu}F^{\mu \nu} +
\frac{\rho(\phi_0)}{4}\; \ep^{\alpha\beta \mu\nu} F_{\a\b} F_{\mu
\nu}. \ee

Clearly, for nearly uniform distributions ($ \phi_0 \approx cte$),
the second term turns out to be a boundary term, and we recover
the standard Maxwell theory without BLS as expected.

It is not difficult to build up a toy model for electromagnetic field
interacting with bosonic spin-0 matter field such that this
theory is obtained by linearization. For example, one could
keep a non-self-interacting electromagnetic
field and encode the non-linearity in the interaction term as in
$L_{ED}[A_{\mu}, \phi]\, =
-{1 \over 4} F_{\mu \nu}F^{\mu \nu} + \frac{\rho(\phi)}{4} \;
\ep^{\alpha\beta \mu\nu} F_{\a\b} F_{\mu \nu} +
L_\phi(\partial_{\alpha}\phi , \phi)$. Further expanding around a solution
$\phi_0(x)$ (and $F^0_{\a\b}$) of the equations of motion\footnote{It
may be even a vacuum solution.} one recovers the free
theory (\ref{Lphi0}).

This approach should be helpful to model electrodynamical phenomena
(e.g. superconductivity) in material media, where
non-homogeneities, as some preferred structure of planes described
by the gradient $\nabla \rho$, are supposed to be important. In
particular, notice that a BLS term would appear in the {\it
interface} between two different material {\it media} if we consider
Goldstone/Higgs-like Lagrangians $L_\phi$, and different vacuum
expectation values in each medium (domain). This proposal will be
explored in more detail in a forthcoming work.


Next, we present another example where electromagnetism is coupled
to vectorial bosonic matter, and exhibits {\it spontaneous}
breaking of Lorentz symmetry\footnote{A close example was given
in Ref. \cite{SBLS}.}. This is a particular case of non-linear electrodynamics
where the background vector $V_\mu$ corresponds to the vacuum configuration
of the field $\phi_\mu$.

  \be  \label{} L_{}[A_{\mu}, \phi_\mu]\,= -{\beta \over 4} F_{\mu
\nu}F^{\mu \nu} +
\frac{1}{4}\phi_{\alpha}A_{\beta}\ep^{\alpha\beta \mu\nu}F_{\mu
\nu} + \frac{\lambda}{4}\, G_{\mu \nu}G^{\mu \nu} + V( \phi_\mu
\phi^\mu ),  \ee where $V$ is a non-linear function and \be
G_{\mu\nu} \equiv \partial_{\mu}\phi_{\nu} - \partial_{\nu}\phi_{\mu}\; .
\ee This is not $U(1)$-invariant but it is Lorentz invariant. When this
symmetry is spontaneously broken, gauge invariance is restored
in the action for the free fields.

In the vacuum state for this field, the kinetic term vanishes
$G^2=0$,
and the maximally symmetric solution corresponds to $G_{\mu\nu}=0$.
Similarly we have
  $F_{\mu\nu}=0$. A local solution of this is

\be \label{vacio}\phi_{0\mu} = \partial_{\mu}\phi_0 \;\; , \;\;
A_{0\mu} = 0 .\ee If $V''|_0 <0$, we have a minimum and
$\phi_{0\mu}(x) = \partial_{\mu}\phi_0 $ must be obtained as a {\it
non-trivial}
solution of the non-linear differential equation
$V'(\,\partial_{\mu}\phi_0 \, \partial^{\mu}\phi_0 \, )=0$.
Substituting these vacuum state configurations in expression
(\ref{1}) we obtain the non-Lorentz invariant (but gauge
invariant\footnote{The $U(1)$ gauge invariance is: $\phi
_{\mu}\to\phi _{\mu}\;\;\; , \; \phi\to\phi \;\;\; , \; A_{\mu}\to
A_{\mu}+
\partial_{\mu}\a $. }) theory: \be L_{}[A_{\mu}, \phi_\mu]\,=
-{\beta \over 4} F_{\mu \nu}F^{\mu \nu} +
\frac{1}{4}(\partial_{\alpha}\phi_0)\, A_{\beta}\ep^{\alpha\beta
\mu\nu}F_{\mu \nu} + \frac{\lambda}{4}\, G_{\mu \nu}G^{\mu \nu} +
\frac12 V''|_0 \, (\phi_\mu-\partial_{\mu}\phi_0)^2 . \ee

\subsection{General action for matter and non-linear coupling
to electromagnetism.}

We consider here a general action for the matter fields with the
standard minimal coupling to the electromagnetic field, but with
an additional non-linear coupling term. We shall argue that when
these sources satisfy an obvious physical requirement (existence
of rest charges), BLS is generated in our approach. The proposed
non-linear coupling is specially important for our purposes, and
it presents another important property related with a non-minimal
coupling, recently worked out with very interesting physical
consequences \cite{todos}.

Let us propose the following action  for fundamental
electrodynamics: \be \label{Ljrest} S_{}= \int dx^4\;\left(
-{\b \over 4} F_{\mu \nu}F^{\mu \nu} + k
(\ep_{\alpha\r\sigma\tau}j^\r F^{\sigma\tau})\; (\ep^{\alpha\beta
\mu\nu} A_\b F_{\mu \nu}) + A_\mu j^\mu \right) + S_{source}[{\bar
\psi}, \psi, \phi, \dots] .\ee Remarkably, this theory satisfies
assumptions I, II and III. In particular, if the minimally coupled
theory $\int dx^4 \,A_\mu j^\mu + S_{source}$ is gauge invariant,
then (\ref{Ljrest}) will satisfy III.

As anticipated, we do not give here any explicit free dynamics for
the sources (which determine $j^\mu$), but we enunciate a very
natural physical condition which $j^\mu$ must satisfy as an additional assumption:

\begin{itemize}

\item[(IV)]({\bf Existence of rest charge}): the action
$S_{source}$ is such that the equations of motion derived from
(\ref{Ljrest}) admit a background solution where the source
is a charge at rest. In other words,
there exists a classical solution where the current is \be
\label{jqt} j^\mu = q \, t^\mu ,\ee where $q$ is a function of the
spatial coordinates.

\end{itemize}

When (\ref{jqt}) is imposed, a
solution $F_0$ of the equations of motion {\it exists}. So,
we get $\ep_{\alpha\r\sigma\tau} j^\r F_0^{\sigma\tau}= q B_\a $,
where $B_\a $ is, by definition, the magnetic field in the rest
frame. Thus, it is clear that when we consider expansion (\ref{1}),
there appears a BLS contribution: \be \label{BLSqk}S_{BLS}=\int d^4x
\; ( qk \; B_\a  )\, \ep^{\alpha\beta \mu\nu} A_\b F_{\mu \nu} ,\ee
where the external vector is proportional to the background magnetic
field, as argued in Section 1 (eq. (\ref{BLS})). As we can see, this
is specially relevant to the study of the Hall effect (next
Section).

Notice, in addition, that another term appears
from expansion (\ref{1}) which remarkably coincides with the
non-minimal coupling introduced in \cite{todos}:\be S_{NM}=\int
d^4x \, V_\a \ep^{\alpha\beta \mu\nu} (j+J)_\b F_{\mu \nu} ,\ee
where $J$ is a current describing a perturbation of the source
(\ref{jqt}). The BLS vector is $V^\a=\ep^{\alpha\beta \mu\nu} A_{0
\b} F_{0\,\mu \nu}$. As we will see in Section 4 (eq. \ref{VB}),
if the non-linear theory is not gauge invariant (but satisfies
III), $V_\mu$ may coincide with the direction of $B_\mu$ even
for a background solution $F_0=$ constant as speculated in
\cite{paschoal}.

As we mentioned above, these non-minimal couplings have been
exhaustively studied also in the context of BLS, but with a $V$ that has
not been generated in the context of the model \cite{todos}. Moreover, this opens
the possibility to explain the non-minimal Pauli-like coupling
\cite{pauli}.

Let us briefly mention that there is another similar theory
with a non-linear coupling, which
is gauge invariant from the beginning (if $S_{source}$ is also):

\be \label{Ljrest-1} S_{}= \int dx^4\;\left( -{\b \over 4} F_{\mu
\nu}F^{\mu \nu} + k (\ep_{\alpha\r\sigma\tau}j^\r F^{\sigma\tau})\;
(\ep^{\alpha\beta \mu\nu} j_\b F_{\mu \nu}) + A_\mu j^\mu \right) +
S_{source}[{\bar \psi}, \psi, \phi, \dots] .\ee By repeating the
above procedure (to expand around solutions (\ref{jqt})), there
appear terms in the lagrangian corresponding precisely to
non-minimal or Pauli coupling (proportional to $B_\a
\ep^{\alpha\beta \mu\nu} J_\b F_{\mu \nu}$). This presents the main
properties of the model proposed in ref. \cite{paschoal} in 2+1
dimensions (in the next Section, it will become more clear how the
planar confinement may appear in a four-dimensional theory of this
type).

\section{Planar electromagnetism and Hall effect}

Quantum Hall effect is the phenomenon of a longitudinal electric
field yielding a transverse current in the presence of an external
perpendicular magnetic field $B_\mu$. The effective physics of this
phenomenon is described by an effective Maxwell-Chern-Simons (or
pure Chern-Simons) theory on a plane perpendicular to $B_\mu$
\cite{zhang,fradkin,FQHE}. However, as discussed in the introduction
of the paper, such description is not clearly well argued from a
four-dimensional viewpoint.

In these models, the vector field is the
so-called ``statistical field" which encodes the effective dynamics
of many charged bodies in the Hall sample and part of the pure
electromagnetic field $A_\mu$. Furthermore, an additional external
electromagnetic field $A^0_\mu$ is considered in order to describe
the field $B_\mu$ \cite{fradkin}.

In the present framework, we are able to reproduce most of this
behavior even at the classical level {\it without} making any
reference to the microscopic structure/dynamics of the charged
matter in the conductor, but based on a possible non-linearity
intrinsic to electrodynamics.

In our approach, the presence of an external
magnetic field automatically implies the appearance of a BLS term
(eq. (\ref{BLS})), and no kind of dimensional reduction is necessary
to obtain a pure Chern-Simons term in the plane of conduction. It
may be obtained from (\ref{BLS}) and, consequently, from a
non-linear theory of electromagnetism, by an elegant and very
simple procedure.

Notice that Ohm's law, $J^i = \sigma^{ij} E_j $ is usually
presented as an effective (constitutive) relation arising from the
collective processes that occur in a conducting medium. They are
used to allow the Maxwell equations to describe the
propagation of electromagnetic fields in material {\it media} and
they do not follow from fundamental electrodynamics.
Remarkably, in the class of theories that we are proposing here, this
relation may be obtained from the equations of motion and it does not
need be imposed by hand.

Let us assume the existence of an uniform background magnetic field
$B^\nu$ in a Hall sample placed in a region $R_H$ of the
spacetime. Consider also a non-linear theory such that, by the
procedure described above, the electromagnetic sector minimally
coupled with an external source reads:\be \label{he1} S_{}=
\int_{R_H} dx^4\;\left(L_{EM}
+ A_\mu J^\mu \right) ,\ee where $L_{EM}$ is given by (\ref{BLS})
(in the previous Section  we proposed a realization of this (see eq.
(\ref{BLSqk}) where the coefficient $\r$ may be identified with
$kq$) ) and $J^\mu$ is the electric current that should be produced
in the Hall sample. This description in which $B_\mu$ is treated as
a fixed background field is actually appropriate, since in the QHE,
typically $|B| \sim$ 1 T.

 Then, the equation of motion derived from this action is: \be
  \frac{\b}{2}
  \partial^\nu F_{\nu\mu} =
\frac12\r(B^2) B^\nu \, \ep_{\mu\nu\a\b}  F^{\a\b} -  J_{ \,\mu}\,
; \ee thus, for a nearly constant electromagnetic field $F$ (or
for low energy),
the LHS of this equation is negligible and we can write
\be \label{CSHE} \frac12\r(B^2) B^\nu \, \ep_{\mu\nu\a\b} F^{\a\b}
=  J_{ \,\mu} .\ee By contracting this equation with $B^\mu$, one
gets $$J_\mu \, B^\mu = 0,$$ which is the condition for planar confinement,
expected to be valid in the context of the Hall
effect.

Considering the spatial components of this equation,
we obtain that the current flux will be deviated by the electric field
according to the equation: \be\label{QHEE} \frac12 \r |B|\,
\ep_{ij} E^j = J_i . \ee
So, the Hall conductance can be expressed in terms of parameters
of the non-linear theory and as a function of $ B_\mu $: \be
\sigma_H = \frac{1}{2} \,|B| \,\rho  ,\ee
and the rest charge density is \be
\label{QHEB} J_0 = \frac12 \r B^\mu b_\mu ,\ee where $b_\mu \equiv
\ep_{\mu\nu\a\b} t^\nu F^{\a\b}$ is the fluctuation of the
magnetic field (i.e, the total magnetic field is $B_\mu + b_\mu
$).

Finally, according to assumption II, viewed as a saddle
point approximation, one finds that, to low energies, the
Chern-Simons theory (\ref{CSHE}) effectively describes the Hall
effect, and $\sigma_H$ results to be quantized in fractions of
$e^2 / h$ in the usual way (see the standard arguments in refs.
\cite{zhang,fradkin,loughlin}). Actually, we can give here a
simple argument for it: let us assume that the magnetic
flux\footnote{Due to the dynamic part of the magnetic field $b$.}
through the total sample area\footnote{As we explained in the end
of Section 2, one can take the region $R_H$ to be $S_H \times (0,
L) \times \Re$ where $S_H$ is the sample plane (two-dimensional),
and $L$ may be arbitrarily small.}, $ \Phi = N_{\Phi}= \int_{S_H}
\vec{b} . d\vec{a} $ is an integer (in units in which $e = \hbar =
1$ \cite{fradkin}). On the other hand, the number of charges must
be $ N e = \int_{S_H} J^0 da $. Thus, substituting these two
expressions into (\ref{QHEB}), and using that $\vec{B} . \vec{b}\,
da = |B| \vec{b} . d \vec{a}$, we get \be N = (\frac12 \r |B| )
N_\phi = \sigma_H N_\Phi \Rightarrow \sigma_H =\frac{N}{N_\Phi}.
\ee So, if each electron is attached to an integer number of flux
units \cite{jain}, $\sigma_H$ is the inverse of this number. In a
forthcoming work we shall analyze this result by considering
quantum aspects of the electric charges and the conductor. The
focus here has been on the description of the behavior of the {\it
electromagnetic field} in this situation.


The above discussion is especially meaningful for the non-linear
models which involve matter fields,
as presented in the previous Section (where $\r$ is given by the
background matter fields), with Ohm's law appearing naturally
as expected and having a natural interpretation. However, let us
point out that an interesting question arises here: ``is it
possible a vacuum contribution to the Ohm conductivity law out of
material media?". If {\it pure} electromagnetism with
$\r(B^2)\neq 0$ exists\footnote{In other words, if there is a classical
solution to the non-linear theory such that the matter fields
vanish and $B_\mu (x) =$ constant $\neq 0 $ in some region $R_H$.},
then, there exists an intrinsic contribution of the
electromagnetic field in itself to Ohm's law and consequently, a
vacuum contribution (due to the electromagnetic background) to the
conductance\footnote{Given a background magnetic field $B_\mu$,
then, in these theories, the external sources $J_\mu$ and the
electromagnetic field $F_{\mu\nu}$ will be related by
(\ref{QHEE}) and (\ref{QHEB})}.

The conventional approach to the quantum Hall effect (QHE) says that it only occurs in
conducting materials. The vector potential is considered as the sum of an effective
field, representing the collective degrees of freedom of many
interacting charged bodies in the conductor, plus the
pure electromagnetic vector potential. In our approach, these degrees of freedom
are encoded in the other fields (here referred to as ``matter
fields") but furthermore a material medium would not be required
to obtain this effect, if pure non-linear theories exist. In the next
section we analyze an example of this situation.

\section{Non-linear toy models.}

In this section, we discuss some examples of non-linear
electrodynamics which exhibit BLS.

First, let us analyze the case of pure electromagnetism, where BLS
can occur without matter fields, due to the presence of an intense
magnetic field. The following example is the most interesting for us since it
satisfies all the assumptions I \-- IV and is similar
to the example discussed in a generic way in Section 1.
We consider the non-linear theory for pure electromagnetism
as being given by:

\be \label{toypuro}{\cal S}_{NL} [A_\mu ] = \int dx^4 \left(
\frac{\beta}{4} F_{\mu \nu}F^{\mu \nu} + \frac{\mu}{4} W_{\mu
\nu\a} W^{\mu \nu\a} \right) ,\ee where \be W = F_{[\mu\nu}
A_{\a]}.\ee Let us consider the following ansatz for a classical solution
\be\label{AVB} A_{0\mu}\equiv \ep_{\mu\nu\a\b} x^\nu B^\a
t^\b + c t_\mu ,\ee where $B_\mu$ and $c$ are constant. We can
directly verify that \be \label{VB} {}^\star W_{0\,\mu} \equiv
{}^\star W_\mu [A_0] = - c\, B_\mu ,\ee where $ {}^\star W_{\b}
\equiv \ep_\b^{\;\;\mu \nu\a} F_{[\mu\nu} A_{\a]} $. To verify
that this is actually a solution of the theory (\ref{toypuro}), we
observe that the equations of motion derived from this action are:
\be
\partial^\mu F_{\mu \nu} \propto A^\a
\partial^\mu W_{\mu\nu\a} \, .\ee Then, (\ref{AVB}) is a  trivial
solution since $F_0, W_0$ are constant. Linearization of this theory
around this solution gives
gauge-dependent terms, Maxwell-like terms (as in the model
(\ref{BLS})), and a BLS contribution given by\be
\label{BLSfinal}{\cal S}_{BLS}
= \int dx^4
\frac{\mu}{2}\,{}^ \star W_{0\, \b} \ep^{\b\mu \nu\a} W_{\mu \nu\a}= -\int
dx^4 \frac{\mu}{2} c \, B_\b \ep^{\b\mu \nu\a} F_{\mu \nu}A_\a , \ee as we
expected.

Notice that the theory defined by (\ref{toypuro}) is not gauge invariant.
Assumption III asserts that linearized electromagnetism about
$F=0$ must be gauge invariant, but this {\it is
not} necessarily true for the full non-linear theory. However, we
can construct a non-linear gauge invariant theory
with a similar solution, assuming a non-linear coupling with matter fields.

A similar non-linear theory of electromagnetism coupled to a
scalar field is \be \label{tiao}{\cal S}_{NL} [{\bar \phi}, \phi , A_\mu ] =
\int dx^4 \left( \frac12 D^\mu {\bar \phi} D_\mu \phi +
\frac{\beta(|\phi|^2)}{4}
F_{\mu \nu}F^{\mu \nu} + \frac{\mu}{4 e^2} {\bar \rho}_{\mu \nu\a} \rho^{\mu
\nu\a} \right) \ee where \be \rho_{\mu\nu\a} = F_{[\mu\nu} D_{\a]}\phi ,
\;\;\; {\bar \rho}_{\mu \nu\a}= F_{[\mu\nu} D_{\a]}{\bar \phi}\ee and
\be\label{covariant} D_\mu \phi = [\partial_\mu - i e A_\mu] \phi \;\; ,\;\;
D_\mu {\bar \phi } =  [\partial_\mu + i e A_\mu]{\bar \phi} .\ee
This theory is invariant under local $U(1)$ gauge transformations:
\be \phi' = e^{i\a}\phi ,\;\;\; {\bar \phi}' = e^{- i\a}{\bar
\phi}, \;\;\; A_\mu = A_\mu - ie \,\partial_\mu \a \ee

As a particular case of (\ref{tiao}), we could also consider
a non-standard scalar field
whose dynamics is given in the non-minimal interaction with the
electromagnetic field, namely the U(1)-invariant theory: \be {\cal
S}_{NL} [{\bar \phi}, \phi , A_\mu ] = \int dx^4
\left(\frac{|\phi|^2}{4} F_{\mu \nu}F^{\mu \nu} + \frac{\mu}{4
e^2}{\bar \rho}_{\mu \nu\a} \rho^{\mu \nu\a} \right) .\ee Notice
that, due to (\ref{covariant}), there clearly appears a term
$\mu\;{\bar \phi} \phi \, W_{\mu \nu\a} W^{\mu \nu\a}$ in the
lagrangian, like in (\ref{toypuro}). A solution similar to
(\ref{AVB}) may be found in this theory but, in this case, the
second term can be eliminated by a gauge transformation and
carried to the $\phi$ sector. It is straightforward to verify that
a background configuration which yields a BLS term
(\ref{BLSfinal}) is: \bea \phi(t) = \zeta e^{i c e t} ,\;\;
\zeta={\rm{constant}} ,\;\; c^2 \mu =1,
\nonumber\\
A_{0 \mu}\equiv \ep_{\mu\nu\a\b} x^\nu B^\a t^\b .\eea This
corresponds to a very special $\phi$ which could
be interpreted as an auxiliary field in order to restore gauge
invariance rather than describing any sort of physical matter.

\section{Remarks and Outline. }

There are many reasons to believe that a fundamental theory should
be relativistic. We have seen that it is possible to get BLS  in
the electromagnetic theory from some relativistic non-linear
theory and to reproduce, consequently, many features of planar electrodynamics.
If electrodynamics is indeed non-linear, many new physical
effects should be expected. In particular, optical activity
(birefringence) is a common feature of electromagnetism with
BLS \cite{astroV2}, which is a general characteristic of the theories we
are considering. Finally, let us remark that the mechanisms proposed
here to obtain BLS emphasizes more the possibility of having BLS in {\it
material media}, where strong background electromagnetic fields or
special charge distributions may occur or may be prepared in
experiments.

This kind of analysis might be performed in more general contexts,
like non-Abelian gauge theories. In particular, it would be
interesting to pursue a similar approach  in gauge formulations of
gravity. Our point of view would be in sharp contrast to the
standard one on this subject, where BLS appears at very large
scales, related to cosmic anisotropies \cite{jac,astroV1,astroV2},
or in the context of superstring physics \cite{kosteleky}. Such
anisotropies of matter at large scale should give place to BLS due
to a non-linear effective electrodynamics in the way we have
proposed here.

\section*{Acknowledgements}

The author thanks  A.L.M.A. Nogueira and Ricardo Paschoal for
fruitful discussions. S. Alves D\'{\i}as and Jos\'e A. Helay\"el
Neto are especially acknowledged for suggestions and criticisms on
the manuscript. CLAF and CNPq are acknowledged for the invaluable
financial help.

\end{document}